\documentclass[draft,journal,a4paper,oneside,onecolumn]{IEEEtran} 
\usepackage{amsmath,amsfonts, amssymb}

\newtheorem{thm}{Theorem}
\newtheorem{cor}[thm]{Corollary}
\newtheorem{lem}{Lemma}
\newtheorem{rem}{Remark}

\newcounter{step}
\newenvironment{algorithm}[2]{%
 \begin{list}{%
   \textsf{Step #2\arabic{step}}}%
  {%
   \usecounter{step}%
   \settowidth{\labelwidth}{\textsf{#1}}%
   \addtolength{\labelwidth}{0mm}%
   \setlength{\leftmargin}{\labelwidth}%
   \setlength{\rightmargin}{0pt}%
   \setlength{\labelsep}{0pt}%
   \setlength{\parsep}{0pt}%
   \setlength{\itemsep}{0pt}%
 }}{\end{list}}

\newcommand{\oA}{\underline{A}}
\newcommand{\bigzerou}{\smash{\lower1ex\hbox{\LARGE $0$}}}
\newcommand{\bigzerol}{\smash{\hbox{\LARGE $0$}}}
\newcommand{\zero}{\boldsymbol{0}}
\newcommand{\B}{\mathcal{B}}
\newcommand{\C}{\mathcal{C}}
\newcommand{\I}{\mathcal{I}}
\newcommand{\J}{\mathcal{J}}
\newcommand{\M}{\mathcal{M}}
\newcommand{\X}{\mathcal{X}}
\newcommand{\Y}{\mathcal{Y}}
\newcommand{\XX}{\boldsymbol{X}}
\newcommand{\YY}{\boldsymbol{Y}}
\newcommand{\MM}{\boldsymbol{M}}
\newcommand{\CC}{\boldsymbol{C}}
\newcommand{\ba}{\boldsymbol{a}}
\newcommand{\xx}{\boldsymbol{x}}
\newcommand{\yy}{\boldsymbol{y}}
\newcommand{\zz}{\boldsymbol{z}}
\newcommand{\cc}{\boldsymbol{c}}
\newcommand{\mm}{\boldsymbol{m}}
\newcommand{\hx}{\widehat{x}}
\newcommand{\hcX}{\widehat{\X}}
\newcommand{\hX}{\widehat{X}}
\newcommand{\tX}{\widetilde{X}}
\newcommand{\tx}{\widetilde{x}}
\newcommand{\oL}{\overline{L}}
\newcommand{\ot}{\overline{t}}
\newcommand{\lrB}[1]{\left[{#1}\right]}
\newcommand{\lrb}[1]{\left\{{#1}\right\}}
\newcommand{\lrsb}[1]{\left({#1}\right)}
\newcommand{\lrbar}[1]{\left|{#1}\right|}
\newcommand{\lrceil}[1]{\left\lceil{#1}\right\rceil}
\newcommand{\fMAP}{f_{\mathrm{MAP}}}
\newcommand{\fMAL}{f_{\mathrm{MAL}}}
\newcommand{\Error}{\mathrm{Error}}
\newcommand{\Risk}{\mathrm{Risk}}
\newcommand{\markov}{\leftrightarrow}
\newcommand{\im}{\mathrm{Im}}
\newcommand{\oH}{\overline{H}}
\newcommand{\uH}{\underline{H}}
\newcommand{\sfB}{\mathsf{B}}

\allowdisplaybreaks

\begin{document}

\title{
  On the Error Probability of\\
  Stochastic Decision and Stochastic Decoding
 \thanks{This paper will be presented in part at the 2017 IEEE International
  Symposium on Information Theory, Aachen, Germany, Jun.\ 25--30, 2017.}
}
\author{
Jun~Muramatsu~\IEEEmembership{Senior Member,~IEEE}
\thanks{J.~Muramatsu is with
 NTT Communication Science Laboratories, NTT Corporation,
 2-4, Hikaridai, Seika-cho, Soraku-gun, Kyoto 619-0237, Japan
 (E-mail: muramatsu.jun@lab.ntt.co.jp).}
and~Shigeki Miyake~\IEEEmembership{Member,~IEEE}
\thanks{S.~Miyake is with
 NTT Network Innovation Laboratories, NTT Corporation,
 Hikarinooka 1-1, Yokosuka-shi, Kanagawa 239-0847, Japan
 (E-mail: miyake.shigeki@lab.ntt.co.jp).
 }
}
  
\maketitle

\begin{abstract}
This paper investigates the error probability of a stochastic decision
and the way in which it differs from the error probability of an optimal
decision, i.e., the maximum a posteriori decision.
This paper calls attention to the fact that
the error probability of a stochastic decision with the a posteriori
distribution is at most twice the error probability
of the maximum a posteriori decision.
It is shown that,
by generating an independent identically distributed random sequence
subject to the a posteriori distribution and making a decision
that maximizes the a posteriori probability over the sequence,
the error probability approaches exponentially the error probability
of the maximum a posteriori decision as the sequence length increases.
Using these ideas as a basis, we can construct stochastic decoders for
source/channel codes.
\end{abstract}
  
\section{Introduction}
This paper considers a decision problem that involves guessing an
invisible state $X$ after observing $Y$, which is correlated with $X$.
In decision theory~\cite[Section 1.5.2]{B85},
an optimal decision rule,
which minimizes the decision error probability,
is called the {\em Bayes decision rule}.

Let $X$ and $Y$ be random variables that take values in
sets $\X$ and $\Y$, respectively,
where we call $X$ a {\em state of nature} or a {\em parameter}
and $Y$ an {\em observation}.
Let $p_{XY}$ be the joint distribution of $(X,Y)$.
Let $p_X$ and $p_Y$ be the marginal distributions
of $X$ and $Y$, respectively.
Let $p_{X|Y}$ be the conditional distribution of $X$ for a given $Y$.
It is well known that
an optimal strategy for guessing the state $X$ consists of finding
$\hx$,
which maximizes the conditional probability $p_{X|Y}(\hx|y)$
depending on a given observation $y$.
Formally, by taking an $\hx$ that maximizes
$p_{X|Y}(\hx|y)$ for each $y\in\Y$,
we can define the function $\fMAP:\Y\to\X$ as
\begin{align}
 \fMAP(y)&
 \equiv\arg\max_{\hx} p_{X|Y}(\hx|y)
 \label{eq:fmap}
 \\
 &=\arg\max_{\hx} p_{XY}(\hx,y),
 \label{eq:ml}
\end{align}
which is a Bayes decision rule.
It should be noted that the discussion throughout this paper
does not depend on choosing states with the same maximum
probability.

When the cardinality $|\X|$ of $\X$ is small,
operations (\ref{eq:fmap}) and (\ref{eq:ml}) are tractable by using
a brute force search.
However, with coding problems, these operations appears to be
intractable because $|\X|$ grows exponentially as the dimension of
$\X$ grows.
In this paper, we assume a situation where operations (\ref{eq:fmap}) and
(\ref{eq:ml}) are intractable.
 
In source coding,
$X$ corresponds to a source output
and $Y$ corresponds to a codeword and side information.
In channel coding, $X$ corresponds to a codeword
and $Y$ corresponds to a channel output,
where the decoding with (\ref{eq:fmap})
is called  {\em maximum a posteriori decoding}.
On the other hand, the decoding method that maximizes
the conditional probability $p_{Y|X}(y|\hx)$ of a channel
is called {\em maximum likelihood decoding}\footnote{We can
interpret the right hand side of (\ref{eq:ml}) as maximizing the likelihood
$p_{XY}(\hx,y)$. For this reason, we might call $\fMAP$
a {\em maximum-likelihood decision rule}.
In a series of papers on the coding problem,
we have called $\fMAP$ a maximum-likelihood decoding
based on this idea.},
which is equivalent to maximum a posteriori decoding
when $X$ is generated subject to the uniform distribution.
In this paper, we call the decision rule with $\fMAP$
the {\em maximum a posteriori decision rule}.

In this paper, we consider a stochastic decision,
where the decision is made randomly subject to a probability
distribution.
We investigate the relationship between the error probabilities
of the stochastic and maximum a posteriori decisions.
To this end, we first investigate general relationships between
the risks of a stochastic decision and optimal decision
and then we apply these results to investigate relationships
between the error probabilities.
Finally, we introduce the construction of stochastic decoders for
source/channel codes.

\section{Definitions of Stochastic Decision}

For a stochastic decision,
we use a random number generator to obtain $\hX\in\X$
after observing $Y$
and let $\hX$ be a decision (guess) about the state $X$.
Formally, we generate $\hX$ subject to the conditional distribution
$q_{\hX|Y}(\cdot|Y)$ on $\X$ depending on an observation $Y$
and let an output be a decision of
$X$, where $X$ and $\hX$ are conditionally independent
for a given $Y$, that is, $X\markov Y\markov\hX$ forms a Markov chain.
The joint distribution $p_{XY\hX}$ of $(X,Y,\hX)$
is given as
\[
 p_{XY\hX}(x,y,\hx)=q_{\hX|Y}(\hx|y)p_{X|Y}(x|y)p_Y(y).
\]
Let us call $q_{\hX|Y}$ a {\em stochastic decision rule}.
As a special case, when $q_{\hX|Y}$
is given by using a function $f:\Y\to\X$ and is defined as
\begin{equation}
 q_{\hX|Y}(\hx|y)
 =
 \begin{cases}
  1 &\text{if}\ \hx=f(y)
  \\
  0 &\text{if}\ \hx\neq f(y),
 \end{cases}
 \label{eq:deterministic}
\end{equation}
we call $q_{\hX|Y}$ or $f$ a {\em deterministic decision rule}.
It should be noted that the maximum a posteriori decision rule is
deterministic.
Let $\chi$ be a support function defined as
\begin{equation}
 \chi(\mathrm{S})
 =
 \begin{cases}
  1 &\text{if a statement $\mathrm{S}$ is true}
  \\
  0 &\text{if a statement $\mathrm{S}$ is false},
 \end{cases}
 \label{eq:support}
\end{equation}
then we have
\begin{equation}
 q_{\hX|Y}(\hx|y)=\chi(\hx=f(y)).
 \label{eq:q=chi}
\end{equation}

\section{Risk of Stochastic Decision}

In this section, we investigate relationships between
the risks of a stochastic decision and optimal decision.
The error probability of a stochastic decision
can be considered as a risk.
We will discuss the error probability of a stochastic decision
in the next section.

Throughout this paper, we assume that $\X$ and $\Y$ are countable sets.
It should be noted that the results do not change when $\Y$ is an
uncountable set,
where the summation should be replaced with the integral.
Let $L$ be a loss function on $\X\times\X$.
We assume that $L$ satisfies some of the following conditions:
\begin{align}
 &L(x,\hx)\geq 0
 \label{eq:L>=0}
 \\
 &L(x,\hx)=0\ \Leftrightarrow\ x=\hx
 \label{eq:L=0}
 \\
 &L(x,\hx)=L(\hx,x)
 \label{eq:L-symmetric}
 \\
 &L(x,\hx)\leq L(x,\tx)+L(\tx,\hx).
 \label{eq:L-subadditive}
\end{align}
It should be noted that
all of above conditions are satisfied when $L$ is a metric.
Furthermore, we assume that
\begin{equation}
 \min_{\hx}\sum_x p_{X|Y}(x|y)L(x,\hx)
 \label{eq:mal}
\end{equation}
exists for a given conditional probability distribution $p_{X|Y}$
and all $y\in\Y$.
This assumption is necessary to consider an optimal decision.
We define
\begin{equation}
 \oL\equiv\sup_{x,\hx}L(x,\hx).
 \label{eq:Lsup}
\end{equation}
Here, we define the {\em risk} $\Risk(q_{\hX|Y})$
of a (stochastic) decision rule $q_{\hX|Y}$ as follows:
\begin{align}
 \Risk(q_{\hX|Y})
 &\equiv
 \sum_y p_Y(y)\sum_{\hx}q_{\hX|Y}(\hx|y)\sum_x p_{X|Y}(x|y) L(x,\hx).
 \label{eq:risk-random}
\end{align}
When $q_{\hX|Y}$ is defined by using $f:\Y\to\X$ and (\ref{eq:deterministic}),
the risk $\Risk(f)$ of a deterministic decision rule $f$ is given as
\begin{align}
 \Risk(f)
 &=
 \sum_{y}p_Y(y)\sum_{x}p_{X|Y}(x|y)L(x,f(y)).
 \label{eq:risk-f}
\end{align}
It should be noted that
the right hand side of this equality
can be derived directly from
(\ref{eq:q=chi}) and (\ref{eq:risk-random}).
That is,
we have $\Risk(f)=\Risk(q_{\hX|Y})$
when $f$ and $q_{\hX|Y}$ satisfy (\ref{eq:deterministic}).

\subsection{Optimal Decision}
Here, we discuss optimal decision rules
which minimize the risk and the error probability.
Let $\fMAL$ be a {\em minimum average loss} decision rule defined as
\begin{equation}
 \fMAL(y)\equiv\arg\min_{\hx}\sum_x p_{X|Y}(x|y)L(x,\hx),
 \label{eq:fmal}
\end{equation}
where we assumed that the minimum on the right hand side
always exists for every $y$.
It should be noted that the discussion throughout this paper
does not depend on choosing states with the same average loss.
We introduce the following well-known lemma.
\begin{lem}
\label{lem:optimal-risk}
Let $(X,Y)$ be a pair consisting of a state $X$ and an observation $Y$
and $p_{XY}$ be the joint distribution of $(X,Y)$.
When we make a stochastic decision with a distribution $q_{\hX|Y}$,
an optimal decision rule minimizing the risk
satisfies $q_{\hX|Y}(\hx|y)=0$ for all $(\hx,y)$ such that
$p_Y(y)>0$ and
 \begin{equation}
 \sum_x p_{X|Y}(x|y)L(x,\hx)>\min_{\hx}\sum_x p_{X|Y}(x|y)L(x,\hx).
 \label{eq:condition-optimal}
 \end{equation}
In particular, the decision rule $\fMAL$ defined by (\ref{eq:fmal})
 minimizes the risk, where $q_{\hX|Y}$ is defined by
(\ref{eq:deterministic}) with $f\equiv\fMAL$.
\end{lem}
\begin{IEEEproof}
From the definition of $\Risk(q_{\hX|Y})$,
we have
\begin{align}
 &
 \Risk(q_{\hX|Y})
 \notag
 \\*
 &
 =
 \sum_y p_Y(y)\sum_{\hx}q_{\hX|Y}(\hx|y)\sum_x p_{X|Y}(x|y)L(x,\hx)
 \notag
 \\
 &
 =
 \sum_y p_Y(y)\sum_x p_{X|Y}(x|y)L(x,\fMAL(y))
 \notag
 \\*
 &\quad
 +      
 \sum_y p_Y(y)\sum_{\hx}q_{\hX|Y}(\hx|y)
 \lrB{\sum_x p_{X|Y}(x|y)L(x,\hx)-\sum_x p_{X|Y}(x|y)L(x,\fMAL(y))}
 \notag
 \\
 &
 =
 \Risk(\fMAL)
 +      
 \sum_y p_Y(y)
 \sum_{\hx}q_{\hX|Y}(\hx|y)
 \lrB{\sum_x p_{X|Y}(x|y)L(x,\hx)-\sum_xp_{X|Y}(x|y)L(x,\fMAL(y))},
 \label{eq:optimal}
 \end{align}
where the last equality
 comes from (\ref{eq:risk-f}).
Since
\begin{align}
 \sum_x p_{X|Y}(x|y)L(x,\hx)-\sum_x p_{X|Y}(x|y)L(x,\fMAL(y))
 &=
 \sum_x p_{X|Y}(x|y)L(x,\hx)-\min_{\hx}\sum_x p_{X|Y}(x|y)L(x,\hx)
 \notag
 \\
 &\geq 0
\end{align}
from the definition of $\fMAL(y)$,
then $\Risk(q_{\hX|Y})$
is minimized only when
$q_{\hX|Y}(x|y)=0$ for all $(x,y)$
satisfying $p_Y(y)>0$ and (\ref{eq:condition-optimal}).
\end{IEEEproof}

\subsection{Upper Bounds of Risk}

In this section, we introduce upper bounds of the risk
of a stochastic decision.
We show the following two lemmas.
\begin{lem}
\label{lem:upperbound-risk-q}
Assume that $L$ satisfies
(\ref{eq:L-subadditive}).
Let $(X,Y)$ be a pair consisting of a state $X$ and an observation $Y$
and $p_{XY}$ be the joint distribution of $(X,Y)$.
When we make a stochastic decision with $q_{\hX|Y}$,
we have
\begin{align}
 \Risk(q_{\hX|Y})
 &\leq
 \Risk(\fMAL)
 +\sum_y p_Y(y)\sum_{\hx} q_{\hX|Y}(\hx|y) L(\fMAL(y),\hx),
 \label{eq:upperbound-risk-q}
\end{align}
where $\fMAL$ is defined by (\ref{eq:fmal}).
\end{lem}
\begin{IEEEproof}
The idea of the following proof comes
from~\cite[Corollary 1 of Theorem 1]{C68}.
We have (\ref{eq:upperbound-risk-q}) as follows:
\begin{align}
 \Risk(q_{\hX|Y})
 &=
 \sum_{y}p_Y(y)
 \sum_{\hx}
 q_{\hX|Y}(\hx|y)
 \sum_{x}
 p_{X|Y}(x|y)
 L(x,\hx)
 \notag
 \\
 &\leq
 \sum_{y}p_Y(y)
 \sum_{\hx}
 q_{\hX|Y}(\hx|y)
 \sum_{x}
 p_{X|Y}(x|y)
 L(x,\fMAL(y))
 \notag
 \\*
 &\quad
 +\sum_{y}p_Y(y)
 \sum_{\hx}
 q_{\hX|Y}(\hx|y)
 \sum_{x}
 p_{X|Y}(x|y)
 L(\fMAL(y),\hx)
 \notag
 \\
 &=
 \sum_{y}p_Y(y)
 \sum_{x}
 p_{X|Y}(x|y)
 L(x,\fMAL(y))
 +\sum_{y}p_Y(y)
 \sum_{\hx}
 q_{\hX|Y}(\hx|y)
 L(\fMAL(y),\hx)
 \notag
 \\
 &=
 \Risk(\fMAL)
 +\sum_{y}p_Y(y)
 \sum_{\hx}
 q_{\hX|Y}(\hx|y)
 L(\fMAL(y),\hx),
\end{align}
 where the inequality comes from (\ref{eq:L-subadditive}).
\end{IEEEproof}

\begin{lem}
\label{lem:upperbound-risk-qoL}
Assume that $L$ satisfies (\ref{eq:L>=0}).
Let $(X,Y)$ be a pair consisting of a state $X$ and an observation $Y$
and $p_{XY}$ be the joint distribution of $(X,Y)$.
When we make a stochastic decision with $q_{\hX|Y}$,
we have
\begin{align}
 \Risk(q_{\hX|Y})
 &\leq
 \Risk(\fMAL)
 +\oL\sum_y p_Y(y)[1-q_{\hX|Y}(\fMAL(y)|y)],
 \label{eq:upperbound-risk-qoL}
\end{align}
where $\oL$ and $\fMAL$ are
 defined by (\ref{eq:Lsup}) and (\ref{eq:fmal}), respectively.
\end{lem}
\begin{IEEEproof}
From (\ref{eq:optimal}),
we have
\begin{align}
 \Risk(q_{\hX|Y})
 &\leq
 \Risk(\fMAL)
 +\sum_{y}p_Y(y)
 \sum_{\hx\neq\fMAL(y)}
 q_{\hX|Y}(\hx|y)\oL
 \notag
 \\
 &=
 \Risk(\fMAL)
 +\oL\sum_{y}p_Y(y)
 [1-q_{X|Y}(\fMAL(y)|y)].
\end{align}
\end{IEEEproof}

\begin{rem}
When we assume that $L$ satisfies (\ref{eq:L=0})
in addition to (\ref{eq:L-subadditive}),
inequality (\ref{eq:upperbound-risk-q}) provides
better bound than (\ref{eq:upperbound-risk-qoL}) because
\begin{align}
 \sum_{y}p_Y(y)
 \sum_{\hx}
 q_{X|Y}(\hx|y)
 L(\fMAL(y),\hx)
 &=
 \sum_{y}p_Y(y)
 \sum_{\hx\neq\fMAL(y)}
 q_{X|Y}(\hx|y)
 L(\fMAL(y),\hx)
 \notag
 \\
 &\leq
 \oL
 \sum_{y}p_Y(y)
 \sum_{\hx\neq\fMAL(y)}
 q_{X|Y}(\hx|y)
 \notag
 \\
 &=
 \oL
 \sum_{y}p_Y(y)
 [1-q_{X|Y}(\fMAL(y)|y)],
\end{align}
where the first equality comes from (\ref{eq:L=0}).
It should be noted that 
we assume (\ref{eq:L>=0})
instead of (\ref{eq:L=0}) and (\ref{eq:L-subadditive})
in Lemma~\ref{lem:upperbound-risk-qoL}.
\end{rem}

\subsection{Stochastic Decision with A Posteriori Distribution}

In this section, we consider the case where
$q_{\hX|Y}(\hx|y)=p_{X|Y}(\hx|y)$ for all $(\hx,y)$, that is,
we make a stochastic decision with the conditional distribution
$p_{X|Y}$ of a state $X$ for a given observation $Y$.
It should be noted that $\hX$ is independent of $X$ for a given $Y$,
where the joint distribution $p_{XY\hX}$ of $(X,Y,\hX)$ is given as
\begin{equation}
 p_{XY\hX}(x,y,\hx)
 =
 p_{X|Y}(\hx|y)p_{X|Y}(x|y)p_Y(y).
 \label{eq:markov-p}
\end{equation}
In the following, we call this type of decision a
{\em stochastic decision with the a posteriori distribution}.
It should be noted that
it may be unnecessary to know (or compute) the distribution $p_{X|Y}$
to make this type of decision.
To make this type of decision,
it is sufficient that we have a random number generator subject to
the distribution $p_{X|Y}(\cdot|y)$ with arbitrary input $y\in\Y$,
where the generated random number is independent of $X$ for a given $y$.

We introduce the following lemma.
\begin{lem}[{\cite[Corollary 1 of Theorem 1]{C68}}]
\label{lem:2risk}
Assume that $L$ satisfies (\ref{eq:L-symmetric}) and (\ref{eq:L-subadditive}).
Let $(X,Y)$ be a pair consisting of a state $X$ and an observation $Y$
and $p_{XY}$ be the joint distribution of $(X,Y)$.
When we make a stochastic decision with $p_{X|Y}$,
the risk of this rule
is at most twice the risk of the decision rule $\fMAL$.
That is, we have
\begin{align}
 \Risk(p_{X|Y})
 &\leq 2\Risk(\fMAL).
 \label{eq:2risk-fmal}
\end{align}
\end{lem}
\begin{IEEEproof}
By letting $q_{\hX|Y}=p_{X|Y}$ and
applying Lemma~\ref{lem:upperbound-risk-q},
we have 
\begin{align}
 \Risk(p_{X|Y})
 &\leq
 \Risk(\fMAL)
 +
 \sum_{y}p_Y(y)
 \sum_{\hx}
 p_{X|Y}(\hx|y)
 L(\fMAL(y),\hx)
 \notag
 \\
 &=
 \Risk(\fMAL)
 +
 \sum_{y}p_Y(y)
 \sum_{x}
 p_{X|Y}(x|y)
 L(x,\fMAL(y))
 \notag
 \\
 &=
 2\Risk(\fMAL)
 \label{eq:proof-random-deterministic}
\end{align}
where the first equality comes from (\ref{eq:L-symmetric}).
\end{IEEEproof}

\subsection{Stochastic Decision by Using Approximated Distribution}

Here, we consider that we use an approximation $q'$ of a
distribution $q$ for a stochastic decision.
Let $d(q,q')$ be the variational distance of
two probability distributions $q$ and $q'$ on the same set as
\begin{align}
 d(q,q')
 &\equiv\frac 12\sum_{x}|q(x)-q'(x)|
 \label{eq:vd}
 \\
 &=\max_{\hcX\subset\X}|q(\hcX)-q'(\hcX)|
 \label{eq:vd-max}
\end{align}
(see~\cite[Eq.~(11.137)]{CT}).
We have the following lemma.
\begin{lem}
\label{lem:approx-risk}
Assume that $L$ satisfies (\ref{eq:L>=0}).
Let $(X,Y)$ be a pair consisting of a state $X$ and an observation $Y$
and $p_{XY}$ be the joint distribution of $(X,Y)$.
When we make decisions with two stochastic decision rules
$q(\cdot|y)$ and $q'(\cdot|y)$ for each $y\in\Y$,
we have
\begin{align}
 \lrbar{\Risk(q)-\Risk(q')}
 &\leq
 \oL d(q\times p_Y,q'\times p_Y),
\label{eq:approx-risk}
\end{align}
where $q\times p_Y(x,y)\equiv q(x|y)p_Y(y)$,
$q'\times p_Y(x,y)\equiv q'(x|y)p_Y(y)$,
and $\oL$ is defined by (\ref{eq:Lsup}).
\end{lem}
\begin{IEEEproof}
Let $\hcX\equiv\lrb{\hx:\ q(\hx|y)\geq q'(\hx|y)}$.
Then we have 
\begin{align}
 \sum_{\hx}
 \lrB{q(\hx|y)-q'(\hx|y)}
 L(x,\hx)
 &=
 \sum_{\hx\in\hcX}
 \lrB{q(\hx|y)-q'(\hx|y)}
 L(x,\hx)
 +
 \sum_{\hx\in\X\setminus\hcX}
 \lrB{q(\hx|y)-q'(\hx|y)}
 L(x,\hx).
\end{align}
Since condition (\ref{eq:L>=0}) implies that $\oL\geq 0$,
then the first term in the right hand side is not less than zero,
and the second term in the right hand side is not greater than zero.
From this fact, we have
\begin{align}
 \sum_{\hx}
 \lrB{q(\hx|y)-q'(\hx|y)}
 L(x,\hx)
 &\leq
 \sum_{\hx\in\hcX}
 \lrB{q(\hx|y)-q'(\hx|y)}
 L(x,\hx)
 \notag
 \\
 &\leq
 \sum_{\hx\in\hcX}
 \lrB{q(\hx|y)-q'(\hx|y)}
 \oL
 \notag
 \\
 &=
 \lrB{q(\hcX|y)-q'(\hcX|y)}
 \oL
 \label{eq:q>=q'}
 \intertext{and}
 \sum_{\hx}
 \lrB{q(\hx|y)-q'(\hx|y)}
 L(x,\hx)
 &\geq
 \sum_{\hx\in\X\setminus\hcX}
 \lrB{q(\hx|y)-q'(\hx|y)}
 L(x,\hx)
 \notag
 \\
 &\geq
 \sum_{\hx\in\X\setminus\hcX}
 \lrB{q(\hx|y)-q'(\hx|y)}
 \oL
 \notag
 \\
 &=
 \lrB{q(\X\setminus\hcX|y)-q'(\X\setminus\hcX|y)}
 \oL.
 \label{eq:q<q'}
\end{align}
Then we have
\begin{align}
 &
 \lrbar{\Risk(q)-\Risk(q')}
 \notag
 \\*
 &=
 \lrbar{
  \sum_{y}p_Y(y)
  \sum_{\hx}
  q(\hx|y)
  \sum_{x}
  p_{X|Y}(x|y)
  L(x,\hx)
  -
  \sum_{y}p_Y(y)
  \sum_{\hx}
  q'(\hx|y)
  \sum_{x}
  p_{X|Y}(x|y)
  L(x,\hx)
 }
 \notag
 \\
 &\leq
 \sum_{y}p_Y(y)
 \sum_{x}
 p_{X|Y}(x|y)
 \lrbar{
  \sum_{\hx}
  \lrB{
   q(\hx|y)-q'(\hx|y)
  }
  L(x,\hx)
 }
 \notag
 \\
 &\leq
 \sum_{y}p_Y(y)
 \sum_{x}
 p_{X|Y}(x|y)
 \max\lrb{
  \lrbar{q(\hcX|y)-q'(\hcX|y)},
  \lrbar{q(\X\setminus\hcX|y)-q'(\X\setminus\hcX|y)}
 }
 \oL
 \notag
 \\
 &\leq
 \oL
 \sum_{y}p_Y(y)
 \sum_{x}
 p_{X|Y}(x|y)
 d(q(\cdot|y),q'(\cdot|y))
 \notag
 \\
 &=
 \oL d(q\times p_Y,q'\times p_Y),
 \label{eq:proof-risk-approx}
\end{align}
where the second inequality comes from (\ref{eq:q>=q'}),
(\ref{eq:q<q'}) and the fact that $\oL\geq0$,
the third inequality comes from the definition (\ref{eq:vd-max}) of
the variational distance,
and the last equality comes from
(\ref{eq:vd}) and the fact that
\begin{align}
 \sum_{y}p_Y(y)
 \sum_{x}
 p_{X|Y}(x|y)
 d(q(\cdot|y),q'(\cdot|y))
 &=
 \frac 12
 \sum_{y}p_Y(y)
 \sum_{x}
 \lrbar{q'(x|y)-q(x|y)}
 \notag
 \\
 &=
 \frac 12
 \sum_{y}
 \sum_{x}
 \lrbar{q'(x|y)p_Y(y)-q(x|y)p_Y(y)}
 \notag
 \\
 &=
 d(q\times p_Y,q'\times p_Y).
 \label{eq:d-prod}
\end{align}
\end{IEEEproof}

\subsection{Stochastic Decision with Random Sequence}
\label{sec:seq-risk}

Here, we assume that a conditional probability $p_{X|Y}$ is
computable. We make a stochastic decision $F(y)$
from a random sequence $\hX^{\ot}\equiv(\hX_1,\ldots,\hX_{\ot})$
as
\begin{align}
 F(y)
 &\equiv \arg\min_{\hx\in\{\hX_t\}_{t=1}^{\ot}}\sum_x p_{X|Y}(x|y)L(x,\hx).
 \label{eq:Fy-risk}
\end{align}
A typical example of a random sequence
$\hX^{\ot}$
is that generated by 
a Markov Chain Monte Carlo method.

We assume that $X$ and $\hX^{\ot}$ are conditionally independent
for a given $Y$, that is,
the joint distribution
$p_{XY\hX^{\ot}}$ of $(X,Y,\hX^{\ot})$ is given as
\begin{equation}
 p_{XY\hX^{\ot}}(x,y,\hx^{\ot})
 =
 q_{\hX^{\ot}|Y}(\hx^{\ot}|y)
 p_{X|Y}(x|y)
 p_{Y}(y),
 \label{eq:sequence}
\end{equation}
where $q_{\hX^{\ot}|Y}(\cdot|y)$ is a joint probability distribution
of $\hX^{\ot}$  for a given $y\in\Y$.
Then, we define a risk $\Risk(F)$ as follows:
\begin{align}
 \Risk(F)
 &\equiv
 E_{XY\hX^{\ot}}\lrB{L(X,F(Y))}
 \notag
 \\
 &=
 E_{Y\hX^{\ot}}\lrB{
  \sum_x p_{X|Y}(x|Y)L(x,F(Y))
 }
 \notag
 \\
 &=
 E_{Y\hX^{\ot}}\lrB{
  \min_{\hx\in\{\hX_t\}_{t=1}^{\ot}}\sum_x p_{X|Y}(x|Y)L(x,\hx)
 }.
 \label{eq:seq-risk}
\end{align}
We have the following theorem.
\begin{thm}
\label{thm:seq-stochastic-risk}
Let $(X,Y)$ be a pair consisting of a state $X$ and an observation $Y$
and $p_{XY}$ be the joint distribution of $(X,Y)$.
When we make a stochastic decision 
$F$ with a random sequence $\hX^{\ot}$ defined
by (\ref{eq:Fy-risk}) and (\ref{eq:sequence}),
the risk $\Risk(F)$ defined by (\ref{eq:seq-risk}) satisfies
\begin{align}
 \Risk(F)
 &
 \leq \min_{t\in\{1,\ldots,\ot\}}
 \Risk(q_{\hX_{t}|Y}),
 \label{eq:seq-risk-bound}
\end{align}
where $q_{\hX_{t}|Y}$ is a conditional marginal distribution
given as 
\begin{equation}
 q_{\hX_t|Y}(\hx_t|y)
 \equiv
 \sum_{(\hx_{t'})_{t'\in\{1,\ldots,\ot\}\setminus\{t\}}}
 q_{\hX^{\ot}|Y}(\hx^{\ot}|y).
 \label{eq:marginal-seq}
\end{equation}
\end{thm}
\begin{IEEEproof}
From (\ref{eq:seq-risk}), we have
\begin{align}
 \Risk(F)
 &=
 E_{Y\hX^{\ot}}\lrB{
  \min_{\hx\in\{\hX_t\}_{t=1}^{\ot}}\sum_x p_{X|Y}(x|Y)L(x,\hx)
 }
 \notag
 \\
 &\leq
 E_{Y\hX^{\ot}}\lrB{
  \sum_x p_{X|Y}(x|Y)L(x,\hX_t)
 }
 \notag
 \\
 &=
 E_{Y\hX_t}\lrB{
  \sum_x p_{X|Y}(x|Y)L(x,\hX_t)
 }
 \notag
 \\
 &=
 \sum_{y}p_Y(y)
 \sum_{\hx_t}q_{\hX_t|Y}(\hx_t|y)
 \sum_x p_{X|Y}(x|y)L(x,\hx_t)
 \notag
 \\
 &=
 \Risk(q_{\hX_t|Y})
\end{align}
for any $t\in\{1,\ldots,\ot\}$,
From this inequality, we have (\ref{eq:seq-risk-bound}).
\end{IEEEproof}

In the following, we assume that
a random sequence $\hX^{\ot}$ is
independent and identically distributed (i.i.d.)
with a distribution $q_{\hX|Y}$ for a given $Y$, that is,
the conditional probability distribution $q_{\hX^{\ot}|Y}$
is given as
\begin{equation}
 q_{\hX^{\ot}|Y}(\hx^{\ot}|y)
 \equiv \prod_{t=1}^{\ot}q_{\hX|Y}(\hx_t|y).
 \label{eq:iid}
\end{equation}
Then we have the following theorem.
From this theorem
and Lemma~\ref{lem:optimal-risk},
we have the fact
that
if
$\oL$ is finite
then
$\Risk(F)$ tends towards $\Risk(\fMAL)$,
where the difference $\Risk(F)-\Risk(\fMAL)$
is exponentially small as the length $\ot$ of a sequence increases.
It should be noted that this theorem includes
Lemma~\ref{lem:upperbound-risk-qoL}
as the case $\ot=1$.
\begin{thm}
\label{thm:seq-iid-risk}
Assume that $L$ satisfies (\ref{eq:L>=0}).
Let $(X,Y)$ be a pair consisting of a state $X$ and an observation $Y$
and $p_{XY}$ be the joint distribution of $(X,Y)$.
When we make a stochastic decision
$F$ with an i.i.d.\ random sequence $\hX^{\ot}$ defined
by (\ref{eq:sequence}) and (\ref{eq:iid}),
$\Risk(F)$
defined by (\ref{eq:seq-risk}) satisfies
\begin{align}
 \Risk(F)
 &\leq
 \Risk(\fMAL)+\oL \sum_{y}p_Y(y)[1-q_{\hX|Y}(\fMAL(y)|y)]^{\ot},
 \label{eq:seq-iid-risk-qoL}
\end{align}
where $\oL$ and $\fMAL$ are
defined by (\ref{eq:Lsup}) and (\ref{eq:fmal}), respectively.
\end{thm}
\begin{IEEEproof}
For a given $y\in\Y$, 
let $\hcX^{\ot}(y)\subset\X^{\ot}$ be defined as
\[
 \hcX^{\ot}(y)\equiv
 \lrb{
 \hx^{\ot}:
 \exists t\in\{1,\ldots,\ot\}\ \text{s.t.}\ \hx_t=\fMAL(y)
 }.
\]
Then we have
\begin{align}
 \Risk(F)
 &=
 E_{Y\hX^{\ot}}\lrB{
  \min_{\hx\in\{\hX_t\}_{t=1}^{\ot}}\sum_x p_{X|Y}(x|Y)L(x,\hx)
 }
 \notag
 \\
 &=
 \sum_y
 p_Y(y)      
 \sum_{\hx^{\ot}\in\hcX^{\ot}(y)}
 q_{\hX^{\ot}|Y}(\hx^{\ot}|y)
 \min_{\hx\in\{\hx_t\}_{t=1}^{\ot}}\sum_x p_{X|Y}(x|y)L(x,\hx)
 \notag
 \\*
 &\quad
 +
 \sum_y
 p_Y(y)      
 \sum_{\hx^{\ot}\notin\hcX^{\ot}(y)}
 q_{\hX^{\ot}|Y}(\hx^{\ot}|y)
 \min_{\hx\in\{\hx_t\}_{t=1}^{\ot}}\sum_x p_{X|Y}(x|y)L(x,\hx)
 \notag
 \\
 &\leq
 \sum_y
 p_Y(y)      
 \sum_{\hx^{\ot}\in\hcX^{\ot}(y)}
 q_{\hX^{\ot}|Y}(\hx^{\ot}|y)
 \sum_x p_{X|Y}(x|y)L(x,\fMAL(y))
 +
 \sum_y
 p_Y(y)      
 \sum_{\hx^{\ot}\notin\hcX^{\ot}(y)}
 q_{\hX^{\ot}|Y}(\hx^{\ot}|y)
 \oL
 \notag
 \\
 &\leq
 \sum_y
 p_Y(y)      
 \sum_x p_{X|Y}(x|y)L(x,\fMAL(y))
 +
 \oL
 \sum_y
 p_Y(y)      
 \sum_{\hx_t\neq\fMAL(y)\ \text{for all}\ t\in\{1,\ldots,\ot\}}
 \prod_{t=1}^{\ot}
 q_{\hX|Y}(\hx_t|y)
 \notag
 \\
 &=
 \Risk(\fMAL)
 +
 \oL
 \sum_y
 p_Y(y)      
 \lrB{
  1-q_{\hX|Y}(\fMAL(y)|y)
 }^{\ot},
 \label{eq:proof-iid}
\end{align}
where the second inequality comes from (\ref{eq:L>=0}) and
the last equality comes from (\ref{eq:risk-f}) and
the fact that
\begin{align}
 \sum_{\hx_t\neq\fMAL(y)\ \text{for all}\ t\in\{1,\ldots,\ot\}}
 \prod_{t=1}^{\ot}
 q_{\hX|Y}(\hx_t|y)
 &=
 \prod_{t=1}^{\ot}
 \lrB{
  \sum_{\hx_t\neq\fMAL(y)}
  q_{\hX|Y}(\hx_t|y)
 }
 \notag
 \\
 &=
 \lrB{1-q_{\hX|Y}(\fMAL(y)|y)}^{\ot}.
 \label{eq:proof-fmalneqhx}
\end{align}
\end{IEEEproof}

\section{Error Probability of Stochastic Decision}

In this section, we apply the results of previous section
to investigate relationships between
the error probabilities of a stochastic decision and an optimal decision.

We have assumed that $\X$ is a countable set.
Then the following lemma guarantees that
the right hand side of (\ref{eq:fmap}) always exists for every $y\in\Y$.
\begin{lem}
Let $q$ be a probability distribution
on a countable set $\X$.
Then the maximum of $q$ on $\X$ always exists, that is,
there is $\hx\in\X$ such that $q(\hx)\geq q(x)$  for any $x\in\X$.
\end{lem}
\begin{IEEEproof}
The lemma is trivial when $\X$ is a finite set.
In the following, we assume that $\X$ is a countable infinite set.

Since $q(x)\leq 1$ for all $x\in\X$,
then $\sup_x q(x)$ always exists, that is,
$q(x')\leq\sup_x q(x)$ for all $x'\in\X$, and
for any $q'<\sup_x q(x)$ there is a $x'\in\X$ such that
$q'\leq q(x')\leq \sup_x q(x)$.

We prove the lemma by contradiction.
Assume that there is no $\hx\in\X$ such that $q(\hx)=\sup_x q(x)$.
Since $\sum_x q(x)=1$ and $q(x)\geq 0$ for all $x\in\X$,
there is $x_0\in\X$ such that $q(x_0)>0$.
From the definition of $\sup_x q(x)$, there is a $x_1\in\X$
such that $q(x_0)\leq q(x_1)<\sup_x q(x)$, where the second
inequality comes from the assumption.
By repeating this argument\footnote{In fact, it is sufficient
to repeat this argument $\lrceil{1+1/q(x_0)}$ times so that
$\lrceil{1+1/q(x_0)}q(x_0)>1$.},
we have a sequence $\{x_i\}_{i=0}^{\infty}$
such that
\[
 0<q(x_0)\leq q(x_1)\leq q(x_2)\leq\cdots<\sup_x q(x).
\]
This implies $\sum_xq(x)\geq\sum_{i=0}^{\infty}q(x_i)=\infty$,
which contradicts $\sum_x q(x)=1$.
\end{IEEEproof}

\begin{rem}
When $\X$ is a finite dimensional Euclidean space,
we can make the same discussion
by quantizing uniformly from $\X$ to a countable set,
where the decision is interpreted
as guessing $X$ with a finite precision.
Then we can apply the results to parameter estimation problems.
\end{rem}

When we consider the error probability of a decision,
we define a loss function as
\begin{equation*}
 L(x,\hx)\equiv \chi(x\neq \hx),
 \label{eq:Lchi}
\end{equation*}
where $\chi$ is defined by (\ref{eq:support}).
It is easily to check that $L$ satisfies
(\ref{eq:L>=0})--(\ref{eq:L-subadditive}) and $\oL=1$.
Then the error probability $\Error(q_{\hX|Y})$ of
a (stochastic) decision rule $q_{\hX|Y}$ is given as
\begin{align}
 \Error(q_{\hX|Y})
 &=
 \sum_{y}p_Y(y)
 \sum_{\hx}
 q_{\hX|Y}(\hx|y)
 \sum_{x}
 p_{X|Y}(x|y)
 \chi(x\neq\hx)
 \notag
 \\
 &=
 \sum_{y}p_Y(y)
 \sum_{x}
 p_{X|Y}(x|y)[1-q_{\hX|Y}(x|y)].
 \label{eq:error-stochastic}
\end{align}
In the last equality,
$1-q_{\hX|Y}(x|y)$ corresponds to the error probability
of the decision rule $q_{\hX|Y}$
after the observation $y\in\Y$,
and $\Error(q_{\hX|Y})$ corresponds to the average of this error probability.
When $q_{\hX|Y}$ is defined by using $f:\Y\to\X$ and (\ref{eq:deterministic}),
the decision error probability $\Error(f)$ of a deterministic decision
rule $f$ is given as
\begin{align}
 \Error(f)
 &=
 \sum_{y}p_Y(y)\sum_{x}p_{X|Y}(x|y)\chi(f(y)\neq x)
 \notag
 \\
 &=
 \sum_{y}p_Y(y)[1-p_{X|Y}(f(y)|y)].
 \label{eq:error-f}
\end{align}
It should be noted that
the right hand side of the first equality
can be derived directly from
(\ref{eq:error-stochastic}) and the fact that
$q_{\hX|Y}(x|y)=\chi(f(y)=x)=1-\chi(f(y)\neq x)$.
That is,
we have $\Error(f)=\Error(q_{\hX|Y})$
when $f$ and $q_{\hX|Y}$ satisfy (\ref{eq:deterministic}).

\subsection{Optimal Decision}

Here, we discuss optimal decision rules
which minimize the error probability.

From the relation
\begin{align}
 \sum_xp_{X|Y}(x|y)L(x,\hx)
 &=
 \sum_xp_{X|Y}(x|y)\chi(x\neq\hx)
 \notag
 \\
 &=1-p_{X|Y}(\hx|y),
 \label{eq:L=chi}
\end{align}
$\fMAL$ is rephrased as
\begin{align}
 \fMAL(y)
 &=\arg\min_{\hx}\sum_xp_{X|Y}(x|y)\chi(x\neq\hx)
 \notag
 \\
 &=\arg\min_{\hx}\lrB{1-p_{X|Y}(\hx|y)}
 \notag
 \\
 &=\arg\max_{\hx}p_{X|Y}(\hx|y)
 \notag
 \\
 &=\fMAP(y).
 \label{eq:fmal=fmap}
\end{align}
Furthermore, condition (\ref{eq:condition-optimal}) is also rephrased as
\[
 1-p_{X|Y}(\hx|y)>\min_{\hx}\lrB{1-p_{X|Y}(x|y)}
\]
which is equivalent to
\begin{equation}
 p_{X|Y}(\hx|y)<\max_{\hx}p_{X|Y}(\hx|y).
 \label{eq:optimal-error}
\end{equation}
From these facts, we have the following well-known lemma.
\begin{lem}
\label{lem:optimal-error}
Let $(X,Y)$ be a pair consisting of a state $X$ and an observation $Y$
and $p_{XY}$ be the joint distribution of $(X,Y)$.
When we make a stochastic decision with a distribution $q_{\hX|Y}$,
an optimal decision rule minimizing the decision error
probability satisfies $q_{\hX|Y}(x|y)=0$ for all $(x,y)$ such that
$p_Y(y)>0$ and
(\ref{eq:optimal-error}).
In particular, the maximum a posteriori decision rule $\fMAP$ defined by
(\ref{eq:fmap})
minimizes the error probability,
where $q_{\hX|Y}$ is defined by $f\equiv\fMAP$ and
(\ref{eq:deterministic}).
\end{lem}

\subsection{Upper Bound of Error Probability}

Here, we introduce an upper bound of the error
probability of a stochastic decision.
We have the following lemma 
from Lemma~\ref{lem:upperbound-risk-qoL}, (\ref{eq:fmal=fmap}),
and the fact that $\oL=1$.
\begin{lem}
\label{lem:upperbound-error-q}
Let $(X,Y)$ be a pair consisting of a state $X$ and an observation $Y$
and $p_{XY}$ be the joint distribution of $(X,Y)$.
When we make a stochastic decision with $q_{\hX|Y}$,
we have
\begin{align}
 \Error(q_{\hX|Y})
 &\leq
 \Error(\fMAP)
 +\sum_y p_Y(y)\lrB{1-q_{\hX|Y}(\fMAP(y)|y)}.
 \label{eq:error-q}
\end{align}
\end{lem}

\subsection{Stochastic Decision with A Posteriori Distribution}

Here, we consider the case where
$q_{\hX|Y}(\hx|y)=p_{X|Y}(\hx|y)$ for all $(\hx,y)$, that is,
we make a stochastic decision with the conditional distribution
$p_{X|Y}$ of a state $X$ for a given observation $Y$.
The joint distribution $p_{XY\hX}$ of $(X,Y,\hX)$ is given by
(\ref{eq:markov-p}).

From Lemma~\ref{lem:2risk} and (\ref{eq:fmal=fmap}), we have
the following lemma, which can also be shown from
Lemma~\ref{lem:upperbound-error-q} and (\ref{eq:error-f}).
In Section~\ref{sec:decoder},
we apply this lemma to an analysis of stochastic decoders
of coding problems.
\begin{lem}[{\cite[Eq.~(29)]{CH67}}]
\label{lem:2error}
Let $(X,Y)$ be a pair consisting of a state $X$ and an observation $Y$
and $p_{XY}$ be the joint distribution of $(X,Y)$.
When we make a stochastic decision with $p_{X|Y}$,
the risk of this rule
is at most twice the risk of the decision rule $\fMAL$.
That is, we have
\begin{align}
 \Error(p_{X|Y})
 &\leq 2\Error(\fMAP).
 \label{eq:2error-fmap}
\end{align}
\end{lem}

Here, we introduce the following inequalities,
which come from
Lemma~\ref{lem:2error}.
In these inequalities,
if either $\Error(\fMAP)$ or $\Error(p_{X|Y})$
vanishes as the dimension (block length) of $X$ goes to infinity,
then the other one also vanishes.
\begin{cor}
\label{cor:bound}
\begin{align*}
 \Error(\fMAP)
 \leq
 \Error(p_{X|Y})
 \leq
 2\Error(\fMAP)
 \\
 \frac 12\Error(p_{X|Y})
 \leq \Error(\fMAP)
 \leq \Error(p_{X|Y}).
\end{align*}
\end{cor}

Here, we introduce another corollary
that comes from Lemmas~\ref{lem:optimal-error} and~\ref{lem:2error}.
\begin{cor}
\label{cor:random}
Let $(X,Y)$ be a pair consisting of a state $X$ and an observation $Y$
and $p_{XY}$ be the joint distribution of $(X,Y)$.
When we make a stochastic decision with $p_{X|Y}$,
the decision error probability of this rule
is at most twice the decision error probability
of {\em any} decision rule $q_{\hX|Y}$.
That is, we have
\begin{equation}
 \Error(p_{X|Y})\leq 2\Error(q_{\hX|Y})
 \quad\text{for any}\ q_{\hX|Y}.
 \label{eq:random}
\end{equation}
\end{cor}
\begin{rem}
Let us consider a situation where $q_{\hX|Y}$ is unknown but
$\Error(q_{\hX|Y})$ can be estimated empirically.
Then the above corollary implies that
the error probability of stochastic decision with the a posteriori
distribution is upper bounded by $2\Error(q_{\hX|Y})$.
For example, when we know empirically that a human being can guess $X$ with
small error probability\footnote{For example, she/he can recognize
handwritten digits with small error probability but
we do not know her/his decision rule $q_{\hX|Y}$ explicitly.},
then the error probability of a stochastic decision with the a posteriori
distribution is also small because it is at most twice the error
probability of her/his decision rule.
\end{rem}

\begin{rem}
Inequality (\ref{eq:random}) is tight in the sense that
there is a pair consisting of $p_{X|Y}$ and $q_{\hX|Y}$ such that
(\ref{eq:random}) is satisfied with equality.
In fact, by assuming that
\begin{align*}
 \X
 &\equiv\{0,1\}
 \\
 p_{X|Y}(0|y)
 &>
 \frac 12
 \\
 p_{X|Y}(1|y)
 &=
 1-p_{X|Y}(0|y)
 \\
 q_{\hX|Y}(0|y)
 &=
 \frac{p_{X|Y}(0|y)^2}{2p_{X|Y}(0|y)-1}
 \\
 q_{\hX|Y}(1|y)
 &=
 1-q_{\hX|Y}(0|y)
\end{align*}
for all $y\in\Y$,
we have
\[
 \Error(p_{X|Y})
 =2\sum_yp_Y(y)p_{X|Y}(0|y)[1-p_{X|Y}(0|y)]
\]
and
\begin{align*}
 2\Error(q_{\hX|Y})
 &=
 2\sum_yp_Y(y)p_{X|Y}(0|y)[1-q_{\hX|Y}(0|y)]
 +2\sum_yp_Y(y)p_{X|Y}(1|y)[1-q_{\hX|Y}(1|y)]
 \notag
 \\
 &=
 2\sum_yp_Y(y)p_{X|Y}(0|y)[1-q_{\hX|Y}(0|y)]
 +2\sum_yp_Y(y)[1-p_{X|Y}(0|y)]q_{\hX|Y}(0|y)
 \notag
 \\
 &=
 2\sum_yp_Y(y)p_{X|Y}(0|y)
 +2\sum_yp_Y(y)[1-2p_{X|Y}(0|y)]q_{\hX|Y}(0|y)
 \notag
 \\
 &=
 2\sum_yp_Y(y)p_{X|Y}(0|y)
 -2\sum_yp_Y(y)p_{X|Y}(0|y)^2
 \notag
 \\
 &=
 2\sum_yp_Y(y)p_{X|Y}(0|y)[1-p_{X|Y}(0|y)]
 \end{align*}
 from (\ref{eq:error-stochastic}).
\end{rem}

\subsection{Stochastic Decision by Using Approximated Distribution}

Here, we consider that we use an approximation $q'$ of a
distribution $q$ for a stochastic decision.
We have the following lemma
from Lemma~\ref{lem:approx-risk} and the fact that $\oL=1$.
It should be noted that the lemma is obtained immediately
from (\ref{eq:vd-max}) by considering the decision error event
measured by using the joint probability distributions $q\times p_Y$ and
$q'\times p_Y$.
\begin{lem}
\label{lem:approx-error}
Let $(X,Y)$ be a pair consisting of a state $X$ and an observation $Y$
and $p_{XY}$ be the joint distribution of $(X,Y)$.
When we make decisions with two stochastic decision rules
$q(\cdot|y)$ and $q'(\cdot|y)$ for each $y\in\Y$,
we have
\begin{align}
 \lrbar{\Error(q)-\Error(q')}
 &\leq d(q\times p_Y,q'\times p_Y),
 \label{eq:approx-error}
\end{align}
where $d$ is defined by (\ref{eq:vd}) or (\ref{eq:vd-max}),
and $q\times p_Y(x,y)\equiv q(x|y)p_Y(y)$,
$q'\times p_Y(x,y)\equiv q'(x|y)p_Y(y)$.
\end{lem}

Applying Lemma~\ref{lem:approx-error}
by letting $q\equiv q_{\hX|Y}$ and $q'\equiv p_{X|Y}$,
we have the following theorem from Lemma~\ref{lem:2error}.
\begin{thm}
Let $(X,Y)$ be a pair consisting of a state $X$ and an observation $Y$
and $p_{XY}$ be the joint distribution of $(X,Y)$.
When we make a stochastic decision with $q_{\hX|Y}$,
the decision error probability
$\Error(q_{\hX|Y})$ is bounded as
\begin{align*}
 \Error(q_{\hX|Y})
 &\leq
 2\Error(\fMAP)+d(q_{\hX|Y}\times p_Y,p_{X|Y}\times p_Y).
\end{align*}
\end{thm}

\subsection{Stochastic Decision with Random Sequence}
\label{sec:seq}

In this section, we assume that a conditional probability $p_{X|Y}$ is
computable. We make a stochastic decision $F(y)$ defined by (\ref{eq:Fy-risk})
from a random sequence $\hX^{\ot}\equiv(\hX_1,\ldots,\hX_{\ot})$.
From (\ref{eq:L=chi}),
we have
\begin{align}
 F(y)
 &=\arg\min_{\hx\in\{\hX_t\}_{t=1}^{\ot}}[1-p_{X|Y}(\hx|y)]
 \notag
 \\
 &=\arg\max_{\hx\in\{\hX_t\}_{t=1}^{\ot}}p_{X|Y}(\hx|y).
 \label{eq:Fy-error}
\end{align}
We
assume that $X$ and $\hX^{\ot}$ are conditionally independent
for a given $Y$, that is,
the joint distribution
$p_{XY\hX^{\ot}}$ of $(X,Y,\hX^{\ot})$ is given as (\ref{eq:sequence}).
From (\ref{eq:seq-risk}) and (\ref{eq:L=chi}), we have
a decision error probability $\Error(F)$ as follows:
\begin{align}
 \Error(F)
 &=
 E_{XY\hX^{\ot}}\lrB{\chi(F(Y)\neq X)}
 \notag
 \\
 &=
 E_{Y\hX^{\ot}}\lrB{
  1-\max_{\hx\in\{\hX_t\}_{t=1}^{\ot}}p_{X|Y}(\hx|Y)
 }.
 \label{eq:seq-error}
\end{align}
We have the following theorem
from Theorem~\ref{thm:seq-stochastic-risk}.
\begin{thm}
\label{thm:seq-stochastic-error}
Let $(X,Y)$ be a pair consisting of a state $X$ and an observation $Y$
and $p_{XY}$ be the joint distribution of $(X,Y)$.
When we make a stochastic decision 
$F$ with a random sequence $\hX^{\ot}$ defined
by (\ref{eq:sequence}) and (\ref{eq:Fy-error}),
the decision error probability $\Error(F)$
defined by (\ref{eq:seq-error}) satisfies
\begin{align}
 \Error(F)
 &
 \leq \min_{t\in\{1,\ldots,\ot\}}
 \Error(q_{\hX_{t}|Y}),
 \label{eq:seq-error-bound}
\end{align}
where $q_{\hX_{t}|Y}$ is a conditional marginal distribution
given as (\ref{eq:marginal-seq}).
\end{thm}

Applying Lemma~\ref{lem:approx-error}
by letting $q\equiv q_{\hX_t|Y}$ and $q'\equiv p_{X|Y}$,
we have the following theorem from Lemma~\ref{lem:2error}.
This theorem implies that if $q_{\hX_t|Y}\times p_Y$ tends towards
$p_{X|Y}\times p_Y$ as $t\to\infty$
(e.g. (\ref{eq:gibbs-approx}) in Appendix)
then the upper bound of error probability $\Error(F)$
is close to at most twice the error probability $\Error(\fMAP)$
of the maximum a posteriori decision.
\begin{thm}
\label{thm:seq-random-approx}
Let $(X,Y)$ be a pair consisting of a state $X$ and an observation $Y$
and $p_{XY}$ be the joint distribution of $(X,Y)$.
When we make a stochastic decision $F$ with a random sequence
$(\hX_1,\ldots,\hX_{\ot})$
defined by (\ref{eq:sequence}) and (\ref{eq:Fy-error}),
the decision error probability
is bounded as
\begin{align*}
 \Error(F)
 &\leq
 2\Error(\fMAP)
 +\min_{t\in\{1,\ldots,\ot\}}
 d(q_{\hX_t|Y}\times p_Y,p_{X|Y}\times p_Y).
\end{align*}
\end{thm}

In the following, we assume that
a random sequence $\hX^{\ot}$ is
independent and identically distributed (i.i.d.)
with a distribution $q_{\hX|Y}$ for a given $Y$, that is,
the conditional probability distribution $q_{\hX^{\ot}|Y}$
is given by (\ref{eq:iid}).
Then we have the following theorem. From this theorem
and Lemma~\ref{lem:optimal-error},
we have the fact that $\Error(F)$ tends towards
the error probability $\Error(\fMAP)$
of the maximum a posteriori decision,
where the difference $\Error(F)-\Error(\fMAP)$
is exponentially small as the length $\ot$ of a sequence increases.
It should be noted that (\ref{eq:seq-iid-error}) includes
Lemma~\ref{lem:upperbound-error-q} as the case $\ot=1$.

\begin{thm}
\label{thm:seq-iid-error}
Let $(X,Y)$ be a pair consisting of a state $X$ and an observation $Y$
and $p_{XY}$ be the joint distribution of $(X,Y)$.
When we make a stochastic decision
$F$ with an i.i.d.\ random sequence $\hX^{\ot}$ defined
by (\ref{eq:sequence}) and (\ref{eq:iid}),
the decision error probability $\Error(F)$
defined by (\ref{eq:seq-error})
satisfies
\begin{align}
 \Error(F)
 &\leq
 \Error(\fMAP)+\sum_{y}p_Y(y)[1-q_{\hX|Y}(\fMAP(y)|y)]^{\ot}.
 \label{eq:seq-iid-error}
\end{align}
In particular, when $q_{\hX|Y}=p_{X|Y}$, we have
\begin{align}
 \Error(F)
 &\leq
 \Error(\fMAP)+\sum_{y}p_Y(y)[1-p_{X|Y}(\fMAP(y)|y)]^{\ot}
 \label{eq:seq-iid-fmap}
 \\
 &\leq
 \Error(\fMAP)+\lrB{1-\inf_{y:p_Y(y)>0}\max_x p_{X|Y}(x|y)}^{\ot}.
 \label{eq:seq-iid-minfmap}
\end{align}
\end{thm}
\begin{IEEEproof}
Inequality (\ref{eq:seq-iid-error}) is shown immediately from
Theorem~\ref{thm:seq-iid-risk}, (\ref{eq:fmal=fmap}), and the fact
that $\oL=1$.
Inequality (\ref{eq:seq-iid-fmap}) is obtained from (\ref{eq:seq-iid-error})
by letting $q_{\hX|Y}=p_{X|Y}$.
Inequality (\ref{eq:seq-iid-minfmap}) is
shown by the fact that
\begin{align}
 \sum_{y}p_Y(y)
 \lrB{
  1-p_{X|Y}(\fMAP(y)|y)
 }^{\ot}
 &\leq
 \sum_{y}p_Y(y)
 \sup_{y:p_Y(y)>0}
 \lrB{
  1-p_{X|Y}(\fMAP(y)|y)
 }^{\ot}
 \notag
 \\
 &=
 \sup_{y:p_Y(y)>0}
 \lrB{
  1-p_{X|Y}(\fMAP(y)|y)
 }^{\ot}
 \notag
 \\
 &=
 \lrB{
  1-\inf_{y:p_Y(y)>0}\max_x p_{X|Y}(x|y)
 }^{\ot}.
\end{align}
\end{IEEEproof}

\begin{rem}
When $|\X|$ is finite, we have
\begin{align}
 \Error(F)
 &\leq
 \Error(\fMAP)
 +\lrB{
  1-\inf_{y:p_Y(y)>0}\max_x p_{X|Y}(x|y)
 }^{\ot}
 \notag
 \\
 &\leq
 \Error(\fMAP)+\lrB{1-\frac1{|\X|}}^{\ot}
 \label{eq:seq-iid-X}
\end{align}
from (\ref{eq:seq-iid-minfmap}), where the last inequality comes from the fact that
\[
 |\X|\max_x p_{X|Y}(x|y)
 \geq \sum_x p_{X|Y}(x|y)
 = 1\ \text{for all}\ y\in\Y
\]
 implies
\[
 \inf_{y:p_Y(y)>0}\max_x p_{X|Y}(x|y)
 \geq \frac1{|\X|}.
\]
On the other hand, we have the same bound (\ref{eq:seq-iid-X})
from (\ref{eq:seq-iid-fmap})
when $q_{\hX|Y}(\cdot|y)$ is the uniform distribution on $\X$
for every $y\in\Y$.
This implies that the stochastic decision with an i.i.d.\ sequence
subject to the a posteriori distribution
is at least as good as that subject to the uniform distribution.
It should be noted that,
when $|\X|$ increases exponentially as the dimension of $\X$ increases,
$\ot$ should also increase exponentially
to ensure that the second term $[1-1/|\X|]^{\ot}$ tends towards zero.
\end{rem}

\section{Construction of Stochastic Decoders}
\label{sec:decoder}

This section introduces applications of the stochastic decision with the
a posteriori distribution to some coding problems.

For simplicity, we assume that
we can obtain an ideal random number subject to a given distribution.
For a given function $A$ on $\X^n$, let $\im A\equiv\{A\xx: \xx\in\X^n\}$.
For a given $\cc\in\im A$, let $\C_A(\cc)\equiv\{\xx: A\xx=\cc\}$.

\subsection{Fixed-length Lossless Compression}
\label{sec:noiseless}

This section introduces a stochastic decoder for
a fixed-length lossless compression
with an arbitrary small decoding error probability.

Let $\mu_{\XX}$ be the probability distribution of $X^n$
and $A:\X^n\to\im A$ be an encoding map, where $\cc\equiv A\xx$
is the codeword of $\xx\in\X^n$.
Then the joint distribution $p_{\XX\CC}$ of a source
$\XX\in\X^n$ and a codeword $\CC\in\im A$ are given as
\[
 p_{\XX\CC}(\xx,\cc)
 =
 \mu_{\XX}(\xx)\chi(A\xx=\cc).
\]
The decoder receives a codeword $\cc$.
By using a stochastic decoder with the distribution
\begin{align}
 p_{\XX|\CC}(\xx|\cc)
 &=
 \frac{\mu_{\XX}(\xx)\chi(A\xx=\cc)}
 {\sum_{\xx}\mu_{\XX}(\xx)\chi(A\xx=\cc)}
 \notag
 \\
 &=
 \frac{\mu_{\XX}(\xx)\chi(A\xx=\cc)}
 {\mu_{\XX}(\C_A(\cc))},
 \label{eq:source-decoder}
\end{align}
we have the bound of error probability from Lemma~\ref{lem:2error}.
This implies that, when we use the encoding map $A$
such that the decoding error probability by using 
the maximum a posteriori decoder vanishes as $n\to\infty$,
the decoding error probability by using the stochastic decoder
with $p_{\XX|\CC}$ also vanishes as $n\to\infty$.
In addition, for a special case of source coding with no side information
at the decoder~\cite{SW2CC},
the fundamental limit $\oH(\XX)$ is achievable with this code,
where $\oH(\XX)$ is the spectrum sup-entropy rate of $\XX$ (see~\cite{HV93}).
It should be noted that the right hand side of (\ref{eq:source-decoder})
is the output distribution of the constrained-random-number generator
introduced in~\cite{CRNG}.

For a given linear code for an additive noise channel,
we can use the constrained-random-number generator
as the stochastic decoder by letting $\XX$ be a channel noise,
$A$ be a parity check matrix, and $\CC$ be the syndrome of $\XX$.
A channel encoder encodes a message to a channel input
$\zz\in\{\zz: A\zz=\zero\}$.
A channel decoder receives a channel output $\yy$,
reproduces a channel noise $\xx=\yy-\zz$
from the syndrome $\cc\equiv A\yy=A[\xx+\zz]=A\xx$
by using the above scheme,
and reproduces a message corresponding to $\zz=\yy-\xx$,
where the decoding is successful when $\xx=\yy-\zz$ is reproduced
correctly from $\cc=A\xx$.
From the above discussion, the decoding error probability is
at most twice the decoding error probability
of the maximal a posteriori decoder.

\subsection{Fixed-length Lossless Compression with Side Information at
  Decoder}

 This section introduces a stochastic decoder for
the fixed-length lossless compression of a source $\XX$ with an
arbitrary small decoding error probability,
where the decoder has access to the side information $\YY$ correlated
with $\XX$.

Let $\mu_{\XX\YY}$ be the joint distribution of $(\XX,\YY)$
and $A:\X^n\to\im A$ be an encoding map, where $\cc\equiv A\xx$
is the codeword of $\xx\in\X^n$.
Then the joint distribution $p_{\XX\YY\CC}$ of 
a source $\XX\in\X^n$, side information source $\YY\in\Y^n$,
and codeword $\CC\in\im A$ is given as
\begin{equation}
 p_{\XX\YY\CC}(\xx,\yy,\cc)
 =
 \mu_{\XX\YY}(\xx,\yy)\chi(A\xx=\cc).
 \label{eq:joint-side-information}
\end{equation}

The decoder receives a codeword $\cc$ and side information $\yy$.
By using a stochastic decoder with the distribution
\begin{align}
 p_{\XX|\YY\CC}(\xx|\yy,\cc)
 &=
 \frac{\mu_{\XX\YY}(\xx,\yy)\chi(A\xx=\cc)}
 {\sum_{\xx}\mu_{\XX\YY}(\xx,\yy)\chi(A\xx=\cc)}
 \notag
 \\
 &=
 \frac{\mu_{\XX|\YY}(\xx|\yy)\chi(A\xx=\cc)}
 {\sum_{\xx}\mu_{\XX|\YY}(\xx|\yy)\chi(A\xx=\cc)}
 \notag
 \\
 &=
 \frac{\mu_{\XX|\YY}(\xx|\yy)\chi(A\xx=\cc)}
 {\mu_{\XX|\YY}(\C_A(\cc)|\yy)},
 \label{eq:side-information-decoder}
\end{align}
we obtain the bound of error probability from Lemma~\ref{lem:2error}.
This implies that, when we use the encoding map $A$
such that the error probability of
the maximum a posteriori decoder vanishes as $n\to\infty$,
the error probability of the stochastic decoder
with $p_{\XX|\YY\CC}$ also vanishes as $n\to\infty$.
In addition, the fundamental limit $\oH(\XX|\YY)$ is achievable with this code~\cite{SW2CC},
where $\oH(\XX|\YY)$ is the spectrum sup-entropy rate of $\XX$
(see~\cite[Theorems 4 and 5]{SV93}).
It should be noted that
the right hand side of (\ref{eq:side-information-decoder})
is the output distribution of the constrained-random-number generator
introduced in~\cite{CRNG}.

\subsection{Channel Coding}

This section introduces a stochastic decoder
for the channel code introduced in~\cite{SW2CC}.

Let $\XX\in\X^n$ and $\YY\in\Y^n$ be random variables
corresponding  to a channel input and a channel output, respectively.
Let $\mu_{\YY|\XX}$ be the conditional probability of the channel
and $\mu_{\XX}$ be the distribution of the channel input.
We consider correlated sources $(\XX,\YY)$
with the joint distribution $\mu_{\YY|\XX}\times\mu_{\XX}$.
Let $A:\X^n\to\im A$ be the source code
with the decoder side information introduced in the previous section.
Let $p_{\XX\YY\CC}$ be the joint distribution defined
by (\ref{eq:joint-side-information}).
The decoder of this source code obtains the reproduction by using the
stochastic decoding with the distribution defined by
(\ref{eq:side-information-decoder}).
Let $\Error(A)$ be the error probability of this source code.
We can assume that
for all $r>\oH(\XX|\YY)$, $\delta>0$ and all sufficiently large $n$
there is a function $A$ such that
\begin{equation}
 \Error(A)\leq \delta,
 \label{eq:errorA}
\end{equation}
where a maximum a posteriori decoder is not assumed for this code.

When constructing a channel code,
we prepare a map $B:\X^n\to\M_n$ and a vector $\cc\in\im A$.
We use the stochastic encoder with the distribution
\begin{align}
 p_{\XX|\CC\MM}(\xx|\cc,\mm)
 &=
 \begin{cases}
  \frac{\mu_{\XX}(\xx)\chi(A\xx=\cc)\chi(B\xx=\mm)}
  {\mu_{\XX}(\C_A(\cc)\cap\C_B(\mm))}
  &\text{if}\ \mu_{\XX}(\C_A(\cc)\cap\C_B(\mm))>0,
  \\
  \text{`encoding error'}
  &
  \text{if}\ \mu_{\XX}(\C_A(\cc)\cap\C_B(\mm))=0
  \end{cases}
  \label{eq:crng-channel}
\end{align}
for a message $\mm\in\M_n$ generated subject
to the uniform distribution on $\M_n$.
It should be noted that the right hand side of the above equality
is the output distribution of the constrained-random-number generator
introduced in~\cite{CRNG}.
The decoder reproduces $\xx\in\X^n$ satisfying $A\xx=\cc$
by using the stochastic decoder with the distribution given by
(\ref{eq:side-information-decoder})
and reproduces a message $B\xx$ by operating $B$ on $\xx$.

In the above channel coding,
let us assume that $r+R<\uH(\XX)$,
\begin{align*}
 r&=\frac1n\log|\im A|
 \\
 R&=\frac1n\log|\im\B|,
\end{align*}
and the balanced-coloring property~\cite{CRNG-VLOSSY}
of an ensemble $(\B,p_{\sfB})$,
where $\uH(\XX)$ is the spectrum inf-entropy rate of $\XX$
and $\M_n\equiv\im\B\equiv\cup_{B\in\B}\im B$.
Then, from \cite[Theorem 1]{SW2CC} and (\ref{eq:errorA}),
we have the fact that
for all $\delta>0$ and all sufficiently large $n$
there are $B\in\B$ and $\cc\in\im A$ such that
the error probability $\Error(A,B,\cc)$ of this channel code
is bounded as
\begin{align}
 \Error(A,B,\cc)
 &\leq \Error(A)+\delta
 \notag
 \\
 &\leq 2\delta.
\end{align}
It should be noted that the channel capacity
\[
 \sup_{\XX}[\uH(\XX)-\oH(\XX|\YY)],
\]
which is derived in~\cite[Lemma 1]{CRNG},
is achievable by letting
$\XX$ be the general source that attains the supremum,
$n\to\infty$, $\delta\to 0$, $r\to\oH(\XX|\YY)$,
and $R\to\uH(\XX)-\oH(\XX|\YY)$.

\subsection{Comments on Stochastic Decoding with Random Sequence}
Here, we comment on the decoding with a random sequence
introduced in Section~\ref{sec:seq}.
For decoding, we can use random sequences
generated by Markov chains (random walks) that converge
to the respective stationary distributions
(\ref{eq:source-decoder}) and (\ref{eq:side-information-decoder}).
In this case, we can apply Theorem~\ref{thm:seq-random-approx}
to guarantee that the decoding error probability
is bounded by
almost
twice the error probability
of the maximum a posteriori decoding
as the sequence becomes longer.
We can also use random sequences
by repeating stochastic decisions independently with the respective
distributions
(\ref{eq:source-decoder}) and (\ref{eq:side-information-decoder}),
where we can use independent Markov chains
to generate an i.i.d.\ random sequence.
In this case, we can use Theorem~\ref{thm:seq-iid-error}
to guarantee that the decoding error probability tends to
the error probability of the maximum a posteriori decoding
as the sequence becomes longer.
When implementing (\ref{eq:Fy-error})
for (\ref{eq:source-decoder}) and (\ref{eq:side-information-decoder}),
it is sufficient to calculate the value of 
the numerator on the right hand side of these qualities
because the denominator does not depend on $\xx$.
The numerator value is easy to calculate
when the base probability distribution is memoryless.

\section{Concluding Remarks}
This paper investigated stochastic decision and stochastic decoding problems.
This paper calls attention to the fact that
the error probability of a stochastic decision with the a posteriori
distribution is at most twice the error probability
of a maximum a posteriori decision.
A stochastic decision with the a posteriori distribution
may be sub-optimal but acceptable when
the error probability of another decision rule
(e.g.\ the maximum a posteriori decision rule) is small.

It is shown that,
by generating an i.i.d. random sequence subject to
the a posteriori distribution and making a decision
that maximizes the a posteriori probability over the sequence,
the error probability approaches exponentially the error probability
of the maximum a posteriori decision as the sequence becomes longer.

When it is difficult to make the maximum a posteriori
decision but the error probability of the decision is small,
we may use the stochastic decision rule with the a posteriori distribution
as an alternative.
In particular, when the error probability of the maximum a posteriori
decoding of source/channel coding tends towards zero as the block length
goes to infinity,
the error probability of the stochastic decoding with the a posteriori
distribution also tends towards zero.
The stochastic decoder with the a posteriori distribution
can be considered to be the constrained-random-number
generator~\cite{CRNG,CRNG-VLOSSY}
implemented by using the Sum-Product algorithm
or the Markov Chain Monte Carlo method (see Appendix).
However, the trade-off between the computational complexity and the
precision of these algorithms is unclear.
It remains a challenge for the future.

\appendix

\section{Algorithms for Constrained-Random-Number Generator}

In this section,
we introduce the algorithms for the constrained-random-number
generator~\cite{CRNG},
which generates random numbers subject to a distribution
\begin{align}
 p_{\XX|\YY\CC}(\xx|\yy,\cc)
 &\equiv
 \frac{p_{\XX|\YY}(\xx|\yy)\chi(A\xx=\cc)}
 {\sum_{\xx}p_{\XX|\YY}(\xx|\yy)\chi(A\xx=\cc)}
 \label{eq:crng}
\end{align}
for a given matrix $A$
and vectors $\cc\in\im A\equiv\{A\xx: \xx\in\X^n\}$, $\yy\in\Y^n$,
where $p_{\XX|\YY}$ is assumed to be memoryless, that is,
there is $\{p_{X_j|Y_j}\}_{j=1}^n$ such that
\[
 p_{\XX|\YY}(\xx|\yy)
 =
 \prod_{j=1}^n p_{X_j|Y_j}(x_j|y_j)
\]
for all $\xx=(x_1,\ldots,x_n)$ and $\yy=(y_1,\ldots,y_n)$.
We review the algorithms introduced in~\cite{CRNG} and~\cite{CRNG-VLOSSY},
which make use of the sum-product algorithm and
the Markov Chain Monte Carlo method, respectively.

In the following algorithms, the symbol `$\leftarrow$' denotes the
substitution.

\subsection{Constrained-Random-Number Generator Using Sum-Product Algorithm}
\label{sec:sum-product}

This section reviews an algorithm that uses the sum-product algorithm.

Let $\{\J_i\}_{i=1}^l$ be a family of subsets of $\{1,\ldots,n\}$.
For a set of local functions $\{f_i:\X^{|\J_i|}\to[0,1]\}_{i=1}^l$,
the sum-product algorithm~\cite{GDL}\cite{KFL01} calculates a
real-valued global function $g$ on $\X$ defined as
\begin{equation*}
  g(x_j)\equiv
  \frac{
    \sum_{\xx\setminus\{x_j\}}
    \prod_{i=1}^lf_i(\xx_{\J_i})
  }{
    \sum_{\xx}\prod_{i=1}^lf_i(\xx_{\J_i})
  }
\end{equation*}
approximately, where the summation $\sum_{\xx\setminus\{x_j\}}$
is taken over all $\xx\in\X^n$ except for the variable $x_j$,
and the function $f_i$ depends only on the
set of variables $\xx_{\J_i}\equiv(x_j)_{j\in\J_i}$.
It should be noted that the algorithm calculates
the global function exactly when the corresponding factor graph has no loop.
Let $\pi_{x_j\to f_i}(x_j)$ and $\sigma_{f_i\to x_j}(x_j)$
be messages calculated as
\begin{align*}
  \pi_{x_j\to f_i}(x_j)
  &\leftarrow
  \prod_{i'\in\{1,\ldots,l\}\setminus\{i\}:j\in\J_{i'}}
  \sigma_{f_{i'}\to x_j}(x_j)
  \\
  \sigma_{f_i\to x_j}(x_j)
  &\leftarrow
  \frac{
    \sum_{\xx_{\J_i\setminus\{j\}}}
    f_i(\xx_{\J_i})
    \prod_{j'\in\J_i\setminus\{j\}}
    \pi_{x_{j'}\to f_i}(x_{j'})
  }{
    \sum_{\xx_{\J_i}}
    f_i(\xx_{\J_i})
    \prod_{j'\in\J_i\setminus\{j\}}
    \pi_{x_{j'}\to f_i}(x_{j'})
  }.
\end{align*}
The summation $\sum_{\xx_{\J}}$ is taken over all $(x_j)_{j\in\J}$,
$\pi_{x_j\to f_i}(x_j)\equiv 1$ when
there is no $i'\in\{1,\ldots,l\}\setminus\{i\}$ such that $j\in\J_{i'}$
and $\sigma_{f_i\to x_j}(x_j)\equiv f_i(x_j)/\sum_{x_j}f_i(x_j)$
when $\J_i=\{j\}$.
The sum-product algorithm is performed by
iterating
the above operations
for every message
$\sigma_{f_i\to x_j}(x_j)$ and $\pi_{x_j\to f_i}(x_j)$
satisfying $j\in\J_i$
and finally calculating the approximation of the global function as
\[
  g(x_j)\approx
  \prod_{i\in\{1,\ldots,l\}:j\in\J_i}
  \sigma_{f_i\to x_j}(x_j),
\]
where we assign initial values to $\pi_{x_j\to f_i}(x_j)$ and
$\sigma_{f_i\to x_j}(x_j)$ when
they appear on the right hand side of the above operations
and are undefined.

In the following, we describe an algorithm for
a constrained-random-number generator.
Let $A\equiv(a_{i,j})$
be an $l\times n$ (sparse) matrix
with a maximum row weight $w$,
where the set $\J_i\equiv\{j\in\{1,\ldots,n\}: a_{i,j}\neq 0\}$
satisfies $|\J_i|\leq w$ for all $i\in\{1,\ldots,l\}$.
Then there is a set $\{\ba_i\}_{i=1}^l$
such that
\[
 A\xx=(\ba_1\cdot\xx_{\J_1},\ba_2\cdot\xx_{\J_2},\ldots,\ba_l\cdot\xx_{\J_l}),
\]
where $\ba_i$ is a $|\J_i|$-dimensional vector and
$\ba_i\cdot\xx_{\J_i}$ denotes the inner product of vectors $\ba_i$ and
$\xx_{\J_i}$.
Let $x_i^j\equiv(x_i,\ldots,x_j)$, where $x_i^j$ is a null string
if $i>j$. Let $\cc\equiv(c_1,\ldots,c_l)\in\X^l$.

\noindent{\bf Constrained-Random-Number Generation Algorithm Using
Sum-Product Algorithm:}
\begin{algorithm}{Step 99}{}
  \item Let $k\leftarrow1$. 
  \item
  Calculate the conditional probability distribution
  $p_{\tX_k|\tX_1^{k-1}Y_1^kC_1^l}$
  defined as
  \begin{align}
   p_{\tX_k|\tX_1^{k-1}Y_1^kC_1^l}(x_k|x_1^{k-1},y_1^k,c_1^l)
   &\equiv
   \frac{\displaystyle
    \sum_{x_{k+1}^n}
    \prod_{j=k}^n
    \mu_{X_j|Y_j}(x_j|y_j)
    \prod_{i=1}^{l}\chi(\ba_i\cdot\xx_{\J_i}=c_i)
   }{\displaystyle
    \sum_{x_k^n}
    \prod_{j=k}^n
    \mu_{X_j|Y_j}(x_j|y_j)
    \prod_{i=1}^{l}\chi(\ba_i\cdot\xx_{\J_i}=c_i)
   },
   \label{eq:sum-product-fk}
  \end{align}
  where $\chi(\cdot)$ is a support function defined by (\ref{eq:support}).
  It should be noted that the sum-product algorithm can be employed
  to obtain (\ref{eq:sum-product-fk}),
  where $\{\mu_{X_j|Y_j}\}_{j=k}^n$ and
  $\{\chi(\ba_i\cdot\xx_{\J_i}=c_i)\}_{i=1}^l$ are local functions,
  and we substitute the generated sequence
  $x_1^{k-1}$ for (\ref{eq:sum-product-fk}).
  If $\chi(\ba_i\cdot\xx_{\J_i}=c_i)$ is a constant
  after the substitution of $x_1^{k-1}$,
  we can record the constant in preparation for the future.
  \item
  Generate and record a random number $x_k$ subject to the
  distribution $p_{\tX_k|\tX_1^{k-1}Y_1^kC_1^l}$.
  \item
  If $k=n$, output $\xx\equiv x_1^n$ and terminate.
  \item
  Let $k\leftarrow k+1$ and go to {\sf Step 2}.
\end{algorithm}

We have the following lemma.
\begin{lem}[{\cite[Theorem 5]{CRNG}}]
\label{lem:sum-product}
Assume that (\ref{eq:sum-product-fk}) is computed exactly.
Then the proposed algorithm generates $\xx\equiv x_1^n$
subject to the probability distribution given by
(\ref{eq:crng}).
\end{lem}

\subsection{Constrained Random Number Generator Using Markov Chain Monte
 Carlo Method}
\label{sec:gibbs}

This section reviews an algorithm that employs
the Gibbs sampling~\cite{H70}, which is a kind of Markov Chain Monte Carlo
method.
In the following algorithm, it is assumed that
$A$ is a systematic matrix
illustrated as
\begin{align*}
A=\lrsb{
\left.
\begin{matrix}
 &&\\
 &\smash{\lower1ex\hbox{\Huge$\oA$}}&\\
 &&
\end{matrix}
\right|
\begin{matrix}
 1&&\bigzerou\\
 &\ddots&\\
 \bigzerol&&1
\end{matrix}
},
\end{align*}
where
the left part $\oA\equiv(a_{i,j})$ is an  $l\times[n-l]$ matrix
and the right part of $A$ is the $l\times l$ identity matrix.
It should be noted that,
when a matrix $A$ is not systematic,
the elementary transformation
and the elimination of redundant rows
can be used to obtain an equivalent condition
represented by a systematic matrix\footnote{It should be noted that
the converted systematic matrix may not be sparse
even when the original matrix is sparse.}.
Let $\ba_j$ be the $j$-th column of $A$
and let $\I_j\equiv\{i: a_{i-n+l,j}\neq 0\}\subset\{n-l+1,\ldots,n\}$.
Let $\kappa$ be the number of iterations.

In the following, we describe an algorithm for
a constrained-random-number generator.
It should be noted that
Steps 2, 5, and 8 realize the stochastic decision defined by
(\ref{eq:Fy-error}), which may be skipped by outputting $\xx$ instead of
$\xx_{\max}$ in Step 9.

\noindent{\bf Constrained-Random-Number Generation Algorithm Using
Markov Chain Monte Carlo Method:}
\begin{algorithm}{Step 99}{}
 \item
 Let $\xx$ be an arbitrary initial sequence satisfying $A\xx=\cc$.
 For example, we can generate $x_1^{n-l}\equiv(x_1,\ldots,x_{n-l})$ randomly
 and let
 \[
  x_i\leftarrow c_{i-n+l}-\sum_{j=1}^{n-l}a_{i-n+l,j}x_j
 \]
 for $i\in\{n-l+1,\ldots n\}$.
 \item
 Let 
 \begin{align*}
  \Lambda&\leftarrow \sum_{j=1}^n\log\mu_{X_j|Y_j}(x_j|y_j)
  \\
  \Lambda_{\max}&\leftarrow \Lambda
  \\
  \xx_{\max}&\leftarrow\xx.
 \end{align*}
 \item
 Let $k\leftarrow 1$.
 \item
 Choose $j\in\{1,\ldots,n-l\}$ uniformly at random.
 \item
 Let 
 \begin{align*}
  \Lambda
  &\leftarrow
  \Lambda - \sum_{i\in\{j\}\cup\I_j}\log\mu_{X_i|Y_i}(x_i|y_i).
 \end{align*} 
 \item
 Calculate the probability distribution $\nu$ defined as
 \begin{align*}
  \nu(x'_j)
  &\equiv
  \frac{
   \mu_{X_j|Y_j}(x'_j|y_j)
   \prod_{i\in\I_j}
   \mu_{X_i|Y_i}(v_{i,j}(x'_j)|y_i)
  }{
   \sum_{x}
   \mu_{X_j|Y_j}(x|y_j)
   \prod_{i\in\I_j}
   \mu_{X_i|Y_i}(v_{i,j}(x)|y_i)
  },
 \end{align*}
 where
 $v_{i,j}(x)\equiv x_i+a_{i-n+l,j}[x_j-x]$
 for $i\in\I_j$.
 \item
 Generate $x'_j$ subject to the probability distribution $\nu$
 and let
 \[
  x_i\leftarrow
  \begin{cases}
   x'_j &\text{if}\ i=j
   \\
   v_{i,j}(x'_j) &\text{if}\ i\in\I_j
   \\
   x_i &\text{otherwise}.
  \end{cases}
 \]
 It should be noted that the renewed sequence $\xx$ satisfies
 $A\xx=\cc$.
 \item
 Let 
 \begin{align*}
  \Lambda
  &\leftarrow
  \Lambda + \sum_{i\in\{j\}\cup\I_j}\log\mu_{X_i|Y_i}(x_i|y_i).
 \end{align*} 
 If $\Lambda<\Lambda_{\max}$,
 let  $\Lambda_{\max}\leftarrow \Lambda$
 and $\xx_{\max}\leftarrow\xx$.
 \item
 If $k=\kappa$, output $\xx_{\max}$ and terminate.
 Otherwise, let $k\leftarrow k+1$ and go to {\sf Step 3}.
\end{algorithm}

We have the following lemma.
\begin{lem}[{\cite[Theorem 5]{CRNG-VLOSSY}}]
\label{lem:gibbs}
For given $(\yy,\cc)$,
let $P_k(\xx|\yy,\cc)$ be the probability of $\xx$
at Step 7 in the above algorithm,
where we can take an arbitrary initial sequence satisfying $A\xx=\cc$
at Step 1.
Then 
\begin{equation}
 \lim_{k\to\infty}d(P_k(\cdot|\yy,\cc),p_{\XX|\YY\CC}(\cdot|\yy,\cc))=0
 \label{eq:stationarylimit}
\end{equation}
for all $(\yy,\cc)$.
\end{lem}

From the above lemma and (\ref{eq:d-prod}), we have
\begin{align}
 \lim_{k\to\infty}d(P_k\times p_{\YY\CC},p_{\XX|\YY\CC}\times p_{\YY\CC})
 &=
 \lim_{k\to\infty}
 \sum_{\yy,\cc}
 p_{\YY\CC}(\yy,\cc)
 d(P_k(\cdot|\yy,\cc),p_{\XX|\YY\CC}(\cdot|\yy,\cc))
 \notag
 \\
 &=
 \sum_{\yy,\cc}
 p_{\YY\CC}(\yy,\cc)
 \lim_{k\to\infty}
 d(P_k(\cdot|\yy,\cc),p_{\XX|\YY\CC}(\cdot|\yy,\cc))
 \notag
 \\
 &=0
 \label{eq:gibbs-approx}
\end{align}
for any $p_{\YY\CC}$.

\end{document}